%% file: iclr2025_conference.tex
\documentclass{article} 
\usepackage{iclr2025_conference,times}

\input{math_commands.tex}

\usepackage{hyperref}
\usepackage{url}
\usepackage{graphicx}

\title{On-Device Watermarking: \\ A Socio-Technical Imperative \\ for Authenticity in the Age of Generative AI}

\author{Houssam Kherraz
\\
Kensho Technologies\\
Cambridge, MA 02138, USA \\
\texttt{houssam.kherraz@kensho.com} \\
}

\iclrfinalcopy 
\begin{document}

\maketitle

\begin{abstract}
As generative AI models produce increasingly realistic output, both academia and industry are focusing on the ability to detect whether an output was generated by an AI model or not. Many of the research efforts and policy discourse are centered around robust watermarking of AI outputs. While plenty of progress has been made, all watermarking and AI detection techniques face severe limitations. In this position paper, we argue that we are adopting the wrong approach, and should instead focus on watermarking via cryptographic signatures trustworthy content rather than AI generated ones. For audio-visual content, in particular, all real content is grounded in the physical world and captured via hardware sensors. This presents a unique opportunity to watermark at the hardware layer, and we lay out a socio-technical framework and draw parallels with HTTPS certification and Blu-Ray verification protocols. While acknowledging implementation challenges, we contend that hardware-based authentication offers a more tractable path forward, particularly from a policy perspective. As generative models approach perceptual indistinguishability, the research community should be wary of being overly optimistic with AI watermarking, and we argue that AI watermarking research efforts are better spent in the text and LLM space, which are ultimately not traceable to a physical sensor.
\end{abstract}

\section{Introduction}

As the capabilities of generative artificial intelligence have advanced in recent years, fear and anxiety have increased with respect to the ability to discern output produced by these AI models. Large Language Models' outputs have gotten so convincing that the Turing test has lost its relevance.
Many universities and professors are scrambling to prevent unauthorized use of these models by their students in their assignments. Despite the large monetary incentive to do this right, and the high number of tools promising the ability to detect AI generated text accurately ~\citep{niuDetectorsEthical}, there are countless stories of students falsely accused of cheating ~\citep{bloombergDetectorsFalsely}, particularly underrepresented students ~\citep{liang2023gptdetectorsbiasednonnative}. \citet{sadasivan2025aigeneratedtextreliablydetected} ran an extensive recursive paraphrasing stress test on a wide range of detection methods, including watermarking, and all methods are nowhere near reliable. Their framework around minimizing Total Variance distance between human and AI text distributions also paint a grim picture around future reliability of such detectors. These detectors have gained wide adoption and caused enough damage with false marketing claims that the FTC looked into it ~\citep{ftcWatchingDetectives}. OpenAI’s classifier was eventually deprecated due to its low accuracy ~\citep{decryptOpenAIQuietly}. This indicates a significant need for these tools, yet models so far have been severely underperforming.
\\
The same holds true for generative images, audio, and video. In popular culture, the Pope Francis generated picture took the internet by storm, and many took it to be real. Deepfakes continue to be a big part of the discourse, especially after Taylor Swift’s deepfake lawsuit. The use of actress Scarlett Johansson's voice by OpenAI without her consent sparked further fears and worries about the impact of audio-visual genAI on society and its understanding of truth and reality~\citep{georgetownOpenAIScarlett}. The potential for misuse, and the ease at which misinformation can be generated and shared, holds a high potential for being very disruptive to our social fabric. While generated text today can most of the time be used as is, the current capabilities for image, audio, and especially video generation are not quite there. Many practitioners are using image generation tools in an iterative process, often using photo editing tools to further polish images. This technological gap provides us with an opportunity to address some issues before they occur at scale.
\\
A major approach in preventing such issues has been via adopting AI watermarking. Surveys of such techniques already exist (see \citet{zhao2024sokwatermarkingaigeneratedcontent} and \citet{boenisch}), so I will not lay them out in this position paper. Some notable ones, from industry research labs especially, include:
\begin{itemize}
    \item \textbf{SyntID} ~\citep{Dathathri2024-kk} - works across text, music, images, and video, and is marketed as an effective solution by Google Deepmind~\citep{deepmindSynthID}. Models need to be built with SynthID incorporated.
\item \textbf{Stable Signature}~\citep{fernandez2023stablesignaturerootingwatermarks} - targets diffusion models. Two models are jointly trained - one that encodes an image and watermark within it, and another that extracts the watermark from augmented versions of the image. The watermark extractor (second model) becomes the AI-generated image detector. 
    \item \textbf{Self-Watermarking through Re-Generation}~\citep{Desu2024GenerativeMA} - works with proprietary black box models as well as open-source ones. It only requires regenerating the output multiple times. The idea is that models fundamentally have some inherent fingerprint rooting that can be leveraged for detection.
\end{itemize}
\section{Fundamental Limitations of AI watermarking
}

While a wide variety of watermarking approaches have been tried so far, empirical results are mixed at best.
\citet{zhang2024watermarkssandimpossibilitystrong} show that under some assumptions on the ability to paraphrase, all text watermarks are removable. \citet{jiang2023evadingwatermarkbaseddetection} show some similar results with image generation - evading watermark detection with postprocessing that is imperceptible to the human eye. \citet{saberi2024robustnessaiimagedetectorsfundamental} show some equivalent findings, further noting how vulnerable watermarking techniques are to spoofing attacks (making real images identified as watermarked ones and hence as AI generated). For the audio modality, the same limitations are faced. \citet{liu2024audiomarkbenchbenchmarkingrobustnessaudio} show that current audio watermarking techniques are very vulnerable and more robust solutions are needed. \cite{liu2024audiomarkbenchbenchmarkingrobustnessaudio} highlight that Synthetic Speech Detectors (SSDs) do not work and are very unprotected against malicious attacks. Note that that detecting synthetic speech is a significantly more constrained problem than detecting synthetic audio overall.

Across modalities, detection of AI generated content is poor and bypassing watermarks is very accessible. Spoofing,
in particular, is very concerning from a policy and societal perspective. If the AI community manages to get the false negative rate high enough and the false positive rate low enough of AI detection techniques, the public could be confident when a piece of content is marked as "AI-generated" that it very likely is. But being able to actively fool systems to confidently tag authentic content as "AI-generated" via spoofing will lead to a complete erosion of the public’s trust in watermarking as a reliable detector of AI generated content (even if outside of spoofing attacks the performance is satisfactory). \\

	To make matters worse, even if our collective research culminates in incredibly reliable watermarking techniques, watermarking will be effectively useless unless the vast majority of AI model providers and users adopt watermarking techniques. From a policy perspective, this will require a large international multilateral agreement from many countries to require watermarking from the AI service providers within their jurisdictions. Even in a world where this monumental feat is achieved, and OpenAI, Hugging Face, and all major providers only host watermarked AI models, we will still collectively face the “genie-out-of-the-bottle” problem. What would effectively stop individuals or organizations from hosting their own models and using them without watermarks? What would deter smaller sovereign jurisdictions from adopting contradictory AI regulatory frameworks to establish competitive advantages that attract AI corporations to their jurisdictions, similar to the Bahamas' approach with crypto companies? As long as there is demand for verisimilar synthetic content, there will be demand for non-watermarked models and bypassing watermark detection. Unless the ML community finds a way to strongly couple watermarking with state-of-the-art model training, such that all state-of-the-art models can only inherently be watermarked, I do not see a promising path forward. At the core, the task for all foundational models, whether in text, audio, image, or video generation, is to sample from a distribution that is as close as possible to the statistical distribution we see in real life. They are trained to imitate reality as closely as possible, and as such, watermarking will, at best, not be a hindrance, and I do not see how it can become coupled with SOTA model training. 

\section{A more bulletproof solution: watermarking audio-visual content at the hardware level via cryptographic signatures
}
\label{gen_inst}

\begin{figure}[h]
\begin{center}
\includegraphics[width=0.85\textwidth]{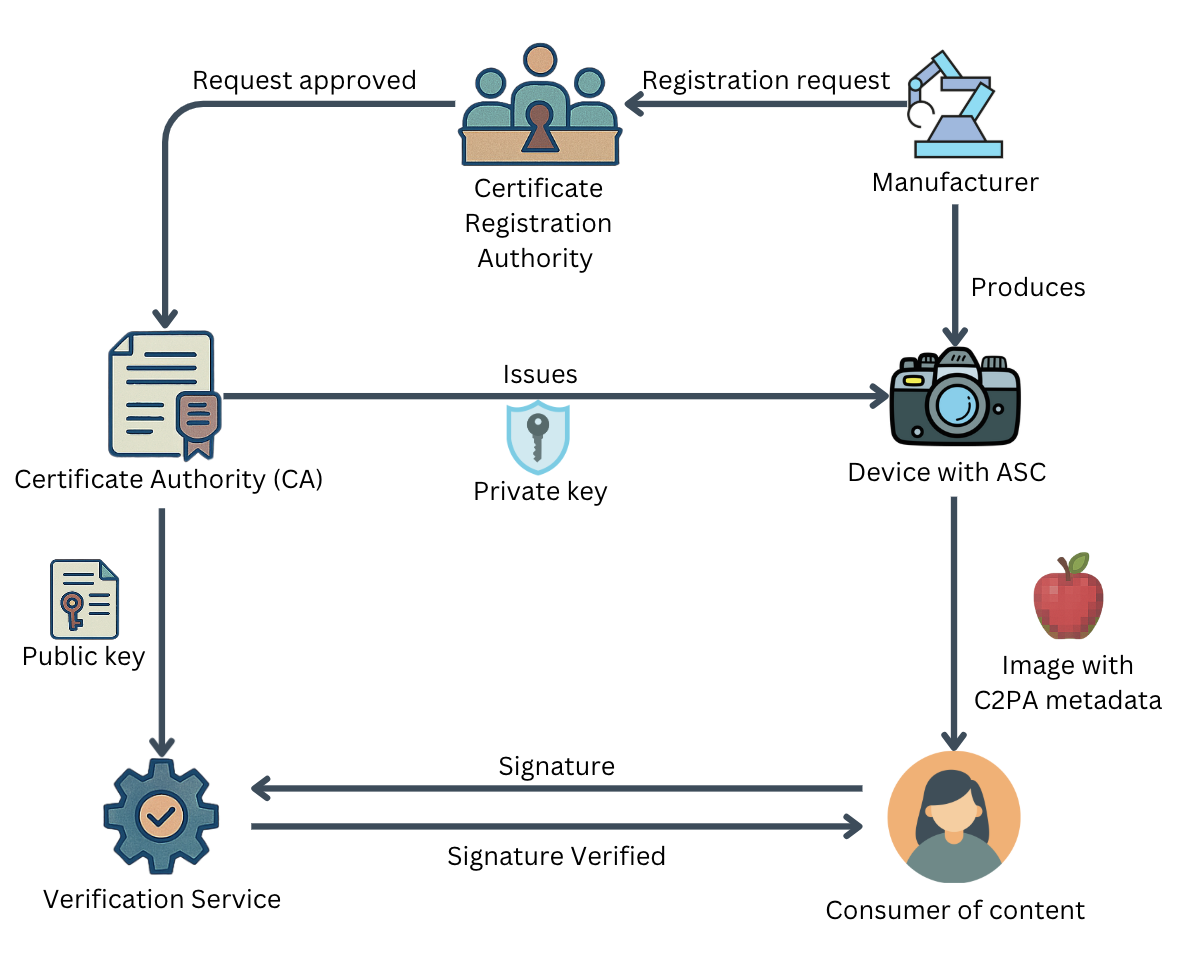}
\end{center}
\caption{Overview of use of CAs for establishing provenance.}
\label{fig:c2pa_ca} 
\end{figure}

An alternative method to the problem of misinformation and authenticity is to approach the issue from the opposite angle: focus on the positive identification of legitimate content.
Instead of watermarking AI content, we watermark “real” content. One notable initiative that currently exists on this front is the Content Authenticity Initiative~\citep{contentauthenticityContentAuthenticity} and the Coalition for Content Provenance and Authenticity (C2PA)~\citep{c2paOverviewC2PA}, an open technical standard to establish trust and the chain of provenance.
\\
The C2PA aims to combat digital misinformation, whether due to AI or not, by establishing a technical standard for tracking the provenance and authenticity of media content, like images, audio and video. The primary method they establish provenance is by making use of Certificate Authorities (CAs) and cryptographic signing, similar to how CAs work for websites and HTTPS. In the context of HTTPS, CAs are trusted entities that issue digital certificates to websites that verify their authenticity and legitimacy. When you go on a website, your browser checks the site’s certificate against a list of trusted CAs, to make sure the website is legitimate and owned by the entity it claims to represent. If the certificate is valid, the browser deems the website as trustworthy and establishes an encrypted connection. C2PA proposes adopting the same framework when viewing images, audio and video. Each piece of content has C2PA provenance data that is cryptographically signed by an entity (the original creator, a later editor, or a publisher). The CAs separately vet and confirm the identities of authors who can attach a signed provenance statement to content via real world diligence. When a consumer of content views it, they can check the provenance data against the CAs to confirm the provenance data and that it has not been tampered with, which is cryptographically guaranteed~\citep{c2paOverviewC2PA}. Figure~\ref{fig:c2pa_ca} provides a high-level overview of this framework. Any alteration to either the content (referred to as the "asset" in the C2PA Specifications) or provenance data would break the mathematical hard binding. Users can confidently view content and see the chain of provenance, that a picture was signed by a journalistic institution they trust and later edited by a news website. Adopting this standard, of course, has many benefits beyond just making sure that content was not AI generated. 
\\

\begin{figure}[h]
\begin{center}
\includegraphics[width=0.85\textwidth]{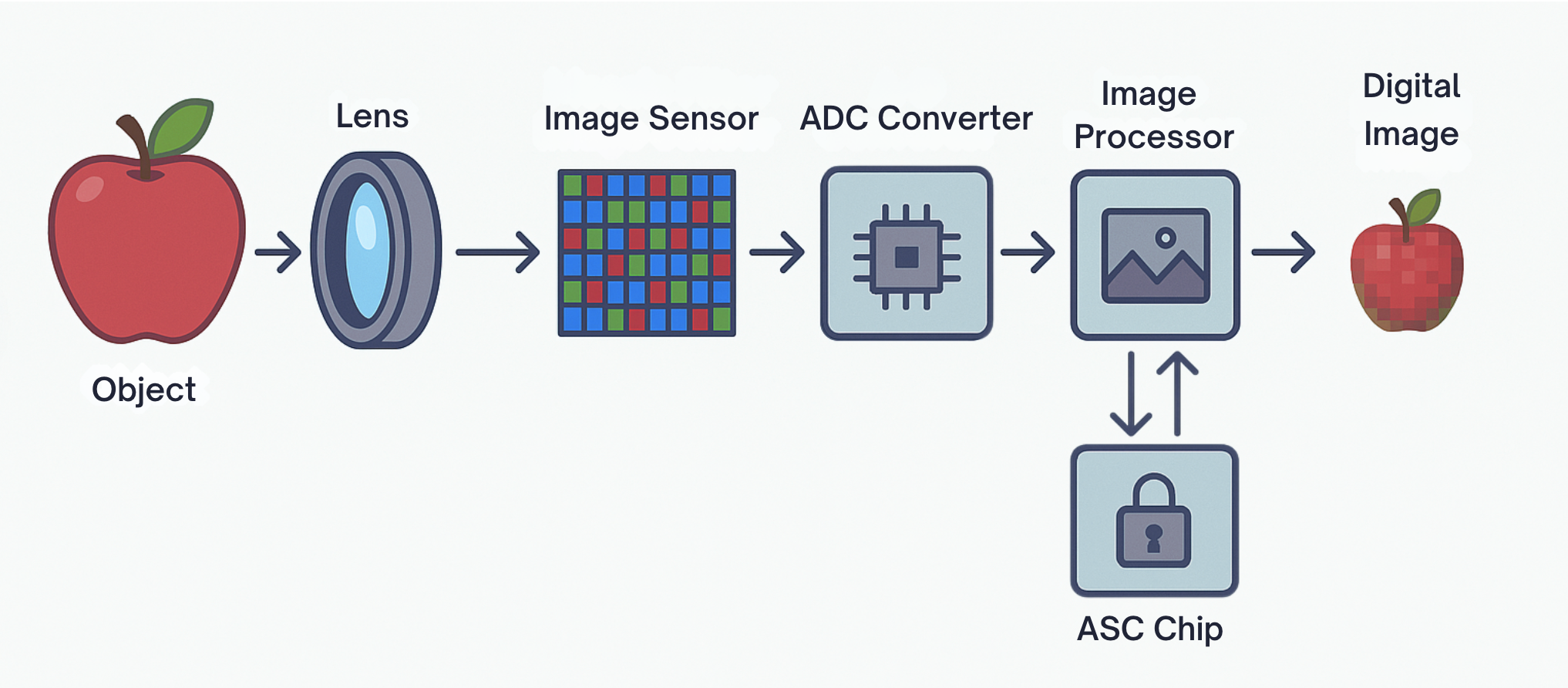}
\end{center}
\caption{Steps involved in capturing an image with a ASC-enabled camera.}
\label{fig:camera_steps} 
\end{figure}

While the C2PA standard lays out a solid foundation, including many aspects that address limitations of watermarking authentic content, it is not aggressive enough in its recommendations and specifications. C2PA ultimately defers the trust verification to the consumer - if the identity of who signed the content is deemed trustworthy by the user then the content is deemed trustworthy. However, this would practically only apply, at best, to governmental and other institutions and to the people who trust them. A government skeptic might believe that an image signed by a governmental institution is still AI generated. Many governments in authoritarian regimes could take advantage of the message of the C2PA standard to spread propaganda. The C2PA standard would also not help establish the trustworthiness of an image or video taken and shared by a complete stranger - e.g someone who did not yet go through the verification process with the C2PA Certification Authorities. As it currently stands, by itself, the C2PA standard is a minor improvement over simply checking the trustworthiness of a host website of content.
\\
Audio, images and video differ from text as a modality in that they are often \textit{grounded in reality}. A lot of the audio, images and video we consume were presumably captured at a moment in time and place, using hardware sensors which translate the analog signal into digital information. This simple fact might prove to be of incredible leverage in fighting against misinformation, the potential harms of AI, and ultimately help improve AI progress. By using existing standards like C2PA and well established cryptographic methods, we can sign audio, images and video at the hardware level. For instance, images are usually captured via a CCD (Charge-Coupled Device) or a CMOS (Complementary Metal-Oxide-Semiconductor) image sensor. The CCD/CMOS sensor turns the photons into electrical charges, which are then converted into bits via an Analog-to-Digital Converter (ADC) after an amplification step. These bits, forming an array, are the original form of the digital image, and are later processed and compressed before being displayed on screens. By introducing an “Authenticity Signature Chip” (ASC) right after the ADC step, the image can be cryptographically signed at the point of creation. Another more pragmatic alternative, is to simply include this within image processors, which are often System-on-Chip devices that include many components. See Figure~\ref{fig:camera_steps} for a summary of the steps described. “Hardware signing” confers a large advantage from a tampering point of view, especially if we also:
\begin{enumerate}
    \item Store a unique cryptographic key for signing content on each ASC. Each camera, microphone or phone can then have a unique key associated with it
    \item Make the ASC design and code proprietary, and its use protected under a license. This would ensure that all manufacturers are taken into account and known to Certificate Authorities (Figure~\ref{fig:c2pa_ca})
    \item Adopt the C2PA standard, to easily integrate with current provenance efforts and avoid reinventing the wheel for enabling modifying the image without losing the original signature and provenance
\end{enumerate}

Microphones also work in a similar fashion where sound is captured by sensors into voltage or capacitance, is amplified and then goes through ADC before a digital processing and encoding step. Similarly, an ASC can be introduced within this pipeline to sign the content at the source, although it will have to occur at the sampling rate to be able to verify a live stream of audio. 
\\
Adopting such a proposal, at a global stage, is not without precedent. Blu-ray Discs (BD) use a combination of encryption, digital rights management (DRM), and licensing agreements to ensure that only licensed and compliant devices can read and play them. To manufacture Blu-ray players, companies must obtain licenses from the Blu-ray Disc Association (BDA). These licensing agreements mandate adherence to the Advanced Access Content System (AACS), BD+, and other security measures. These licenses are necessary to access the decryption keys required to play protected content. The system assigns device-specific keys, and if a key is compromised (e.g., due to hacking), the AACS licensing authority can revoke it in future Blu-ray releases, preventing non-compliant or hacked players from accessing new discs. Further measures are in place to prevent unauthorized duplication and such. These mechanisms collectively ensure that only licensed devices can read Blu-ray Discs while making piracy and unauthorized use more difficult.
\\
If the BDA managed to build an entire ecosystem of BluRay players and producers that adhere to such a system, I do not see why doing so for watermarking audio, images and videos at the hardware level would not be feasible. Policy makers would have a precedent and existing playbook to follow. Enforcing policy on hardware manufacturers is also much simpler, with clear precedents and blueprints to follow, than enforcing policy on every AI model on the planet. Manufacturers tend to be fewer in number, a lot more concentrated in specific geographical areas and jurisdictions, thus hardware production is easier to control and regulate. Given the official stance of both many big tech companies (e.g Google’s commitment to promote trustworthy information~\citep{blogCommitmentAdvancing}) and governments (e.g China banning genAI without watermarks~\citep{arstechnicaChinaBans} and the Biden-Harris AI Executive Order~\citep{bidenharrisai}), passing policy that is more foolproof and helps resolve other non-AI related spread of misinformation should be an easier task. Further, many camera and microphone manufacturers will be financially incentivised to agree on such a standard. As generative audio, images and video improves, there will be less demand for these devices: why take a photo when you can generate the perfect one? Why buy a high quality camera when an AI model can improve the quality of pictures taken by a low quality one? Why buy a high-end microphone for podcasting when a voice clone can just read your script better than you ever will? These financial pressures will push audio-visual device manufacturers to seek a competitive advantage, and they will find this competitive advantage by giving their customers the ability to claim provable authenticity. 
\\
Many positive externalities come from adopting this hardware-enabled cryptographic signing of original content rather than watermarking AI content:
\begin{itemize}
    \item \textbf{This not only helps with fake AI content, it also helps with fake human-made content}. Because of the provenance data, photoshopped or altered content becomes verifiable as well, and people can find and check the original content. Further, if other metadata is encoded like time and GPS coordinates, other types of misinformation can also be curbed. A lot of current “fake”/misleading pictures that spread are taken at a different context but reused in a new one to spark outrage, controversy and clicks (e.g horrible pictures from a past conflict reused to spread fear and hate in the context of a new conflict)
    \item \textbf{AI researchers can focus on advancing audio-visual generative models, without worrying about misinformation impacts}. There will be less reasons to close-source, which is a net benefit to the community. These tools would no longer be viewed as dangerous if there is a reliable path to identify content captured in the real world.
\end{itemize}

\section{Limitations}
\label{headings}

The approach of cryptographic watermarking of original content at the hardware level comes with its own set of challenges, although more tractable than AI watermarking.

\subsection{Modifications and Edits}

One primary issue is that a lot of audio-visual content is usually processed after capture. People almost never actually consume the original content - images are usually edited for lighting or resized, audio is also edited for brevity, noise removal and other reasons. These edits will require users to view the original content at the head of the provenance and compare it with the final product consumed, since any edit of the original asset will show a mismatch with the original signature. In fact, many of these edits are likely to occur on the device itself. For example, cameras today do a lot of software processing of captured images – even a RAW image is almost never the direct result of the simple pipeline described in Section 3. The lack of validity of the cryptographic signature due to any edits, even minor, is by design. One step along the chain of edits could easily involve a major modification, including changing the image or audio entirely with a generated one as the edit. This presents an interesting Human-Computer Interaction challenge: how to make sure enough people check provenance metadata and the original asset such that dishonest actors are disincentivized to provide an AI generated asset as the last edit in a C2PA provenance chain? Well-designed human-centered software can solve for this by making the UX simple enough to check. If C2PA is adopted more broadly, or regulation is passed to adopt it, browsers and other applications (being the primary vehicles through which people consume audio-visual content) would include features to further enable this: preloading the original asset so the user can quickly compare by double tapping or hovering on the image or audio, enabling any user to easily flag the asset as dubious to take it down and retract the C2PA certificate of the final signer as valid, incentivizing websites to check the provenance of their assets by keeping score of those that don’t and tagging them as untrustworthy, etc. The data is all present, and incentives can be aligned to make sure the original asset and signatures are checked.
While this adds a layer of complexity, consumers of digital media will become increasingly aware of the edits to raw audio-visual content, beyond AI. Unrealistic beauty standards and major photoshopping will become more obvious in pictures and social media. Ironically, if hardware crypto-watermarking and C2PA become a widely used standard, the current challenge posed by AI generation might possibly usher in a new era of authenticity in the future.

\subsection{Privacy and Freedom}

\begin{figure}[h]
\begin{center}
\includegraphics[width=0.85\textwidth]{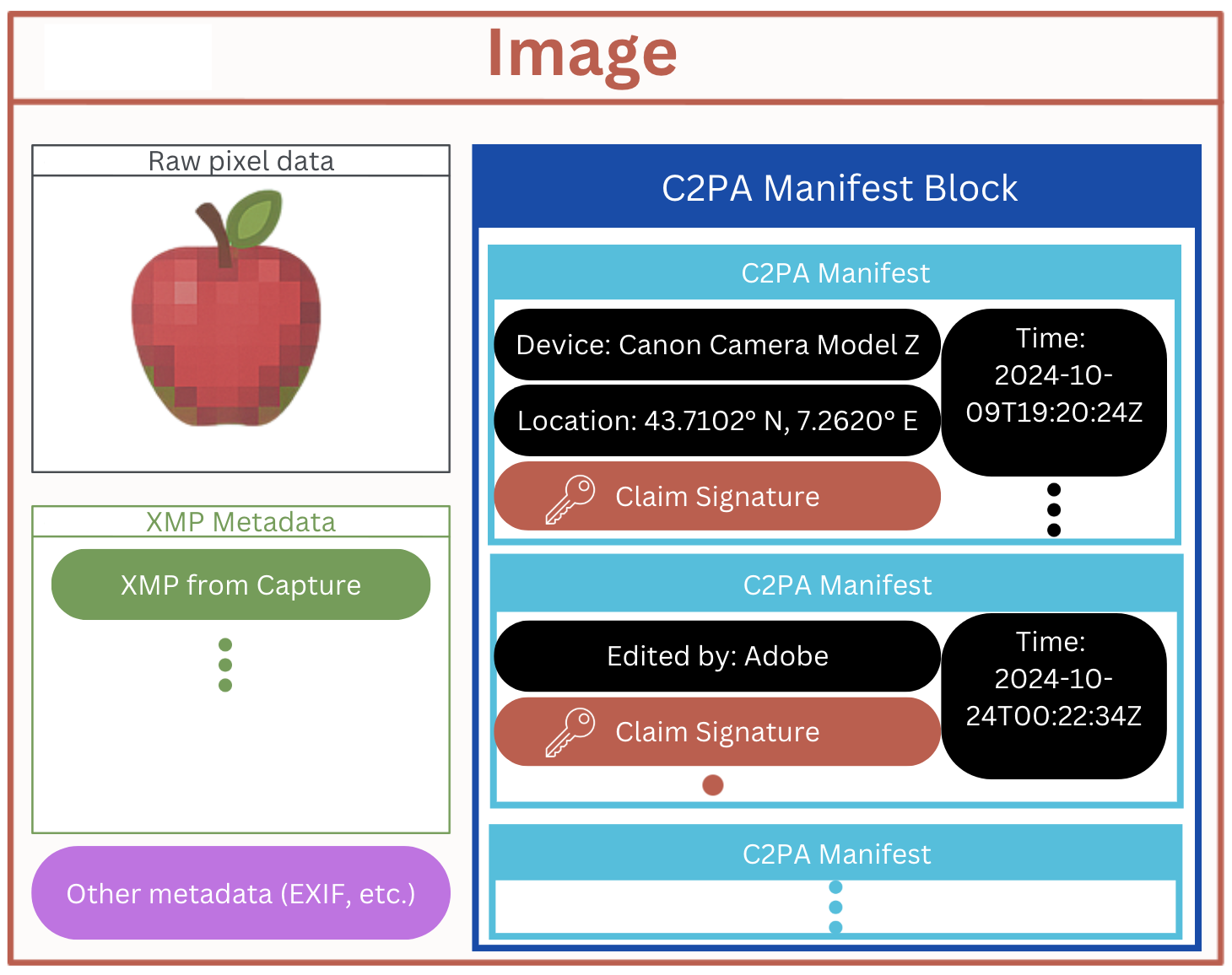}
\end{center}
\caption{Example breakdown of a C2PA asset}
\label{fig:c2pa_image_breakdown} 
\end{figure}

Another major concern with using hardware cryptographic signatures is around privacy and freedom. Many people are already concerned that phones have unique identifiers (International Mobile Equipment Identity, MAC addresses, Advertising IDs and others), and how easy it can be to match a person’s identity with their online activity due to these unique identifiers. 
\\
 Given that the ASC will have a unique key, people could identify who took a picture based on previous pictures taken by the same camera. An actor with some resources can scrape the web for all pictures using the same cryptographic key and use them to find hints and triangulate on the photographer’s identity. This lack of privacy could potentially endanger journalists who are wanted by certain regimes, and prevent great journalism from seeing the light of day due to fears around safety. Similar fears exist around signing metadata content as well, like time and place, and adding it to the provenance chain, but those can be removed with the C2PA redaction process. For a high level overview of what is typically included as metadata in a C2PA asset, see Figure~\ref{fig:c2pa_image_breakdown}.\footnote{You can read more about what is included in a C2PA Manifest in https://opensource.contentauthenticity.org/docs/manifest/manifest-examples/}
\\

A few solutions exist for this problem with different sets of tradeoffs. One potential solution is to have a lot of devices share the same key. This protects individual privacy, since the key is no longer Personally Identifiable Information (PII), but introduces another challenge: if a key is hacked or a device is reverse engineered to sign arbitrary content, a lot of devices and a lot of content will become untrustworthy, since many copies of that key exist and are actively being used. With unique keys, the damage is contained to a singular device and a lot less content, and it becomes easier to retract the key as untrustworthy in provenance checks. Another solution, with arguably a better set of tradeoffs, is to give each ASC chip a set of unique keys to use rather than a singular one. Each chip then randomly picks a key to sign with every time. The mapping from key to device is still unique, which limits damage if a key is compromised, while people’s privacies are better protected. The computational complexity, on the other hand, only increases slightly, making it a beneficial tradeoff.

\subsection{The read-but-not-remove problem}

Based on the current C2PA specifications, the watermarking/signing scheme proposed in this paper suffers from removal attacks. Like \citet{cox:inria-00504528} states: the capability to read-but-not-remove is almost a holy grail of digital watermarking. However, the robustness guarantees in \citet{Jiang2024CertifiablyRI} makes us optimistic and could be the basis of future work. \\
Ultimately, the \textit{removal} of the watermark in our case has a limited impact. Unlike with AI watermarking, where a removal attack leads to the AI generated content to be considered \textit{real}, in our case, an \textit{authentic} asset will simply not have the provable chain that it was originally captured on a piece of hardware. The capability to \textit{prove} authenticity remains valid, which is what matters in establishing trust in our framework. Content without a cryptographically signed provenance chain \textit{may} or \textit{may not} be \textit{authentic} - since many devices (like the currently manufactured ones) will not be cryptographically signing their outputs.

\subsection{Taking a picture of a generated picture or recording a generated recording}
Finally, a potential major flaw in this approach is that people could simply generate an image using a diffusion model and then take a picture of it with their camera (whether directly on screen, after printing it or projecting it). This might allow nefarious actors to claim authenticity of AI generated content. Similar scenarios can be imagined with video or audio, via recording AI generated media. One potential solution for images and videos is to embed depth data into the provenance metadata. Most modern cameras and phones come with depth sensors, whether it is through stereo vision (dual cameras taking the same picture to figure out distance), Infrared pulse sensors or LiDAR (Light Detection and Ranging), and that data could be embedded to further demonstrate authenticity. An alternative solution is to train AI models to detect these specific cases, which presents a significantly narrower and more tractable problem compared to identifying all AI-generated content. Projections, prints, screens and recordings of audio played via speakers, all potentially have inherent artifacts that AI models could learn, even when imperceptible to the human eye. This task is probably a more worthwhile use of the collective talent of the AI community than trying to solve the AI watermarking problem, which, as discussed earlier, will not solve AI misinformation issues.

\subsection{What about the text modality?}
The paper so far discusses solutions for audio, image and video modalities. Text, unfortunately, is not grounded in the physical world. Hardware sensors do not produce text. Words are ultimately the product of our own imagination and minds. C2PA can still apply to documents, with original human authors with unique keys marked as the original producers, but those original authors could still use genAI to produce them. This could at the very least help prevent the vast spread of bot generated content at scale. Search engines will be incentivized to prioritize articles and documents with valid provenance data (associated with a verified individual or organization) in search results, as these will ultimately lead to more engagement and profit. 
Ultimately, LLM watermarking continues to be a great area of potential impact for AI watermarking research.

\section{Conclusion}

Watermarking AI-generated content has been a primary approach to addressing concerns about misinformation and authenticity, but as a lot of literature has shown, fundamental limitations make it an unreliable long-term solution. Although watermarking may provide some immediate safeguards, it is inherently vulnerable to removal, evasion, and spoofing attacks. Furthermore, even if universal adoption was achieved among major AI providers, there would still be demand for non-watermarked models, making enforcement nearly impossible. Given these challenges, a more promising approach is to shift the focus from marking AI-generated content to verifying real content at the hardware level.
\\
By embedding cryptographic signing mechanisms directly into image sensors, microphones, and other hardware components, we can create an ecosystem where authenticity is provable from the moment of capture. The C2PA standard provides a strong foundation for this, and lessons from industries such as Blu-ray DRM enforcement suggest that such an approach is feasible. This method not only helps combat AI-generated misinformation, but also strengthens trust in all digital content, including manipulated human-made content. Additionally, it allows AI researchers to focus on advancing generative models without exacerbating fears around misinformation, and gives policymakers a more structured path forward.
\\
However, privacy concerns, the potential for circumvention via re-recording, and the need for broad industry adoption all present hurdles that must be addressed. For the text modality, which lacks physical grounding, traditional AI watermarking approaches may still have merit and deserve continued research attention. Despite these limitations, cryptographically signing audio-visual content at the hardware level offers a more practical and robust solution to the authenticity problem than watermarking AI content.

\subsubsection*{Acknowledgments}
I am very grateful to my manager, Diana Mingels, for her continued support and encouragement in my career. Also, I thank my employer, Kensho Technologies (an S\&P Global company), for their backing and fostering a culture of curiosity and innovation that made this work possible. I would also like to thank my friend, Ebrahim Al Johani, a senior photonics engineer, for his help in proofreading this manuscript and for double checking the hardware feasibility of the proposed approach. \\
Finally, I would like to give a shoutout to my team (the ML Ops team at Kensho: Alec Alameddine, Brian Chen, Ilya Yudkovich, Justin Tervala, Lance Luo, Mahyar Raza, Matthew Theisen and Mike Arov) for being fantastic colleagues to work with in my day-to-day work (unrelated to this research work).

\bibliography{iclr2025_conference}
\bibliographystyle{iclr2025_conference}


\end{document}

%% file: math_commands.tex
\usepackage{amsmath,amsfonts,bm}

\def\eqref#1{equation~\ref{#1}}

\def\1{\bm{1}}

\DeclareMathAlphabet{\mathsfit}{\encodingdefault}{\sfdefault}{m}{sl}
\SetMathAlphabet{\mathsfit}{bold}{\encodingdefault}{\sfdefault}{bx}{n}

